\documentclass[%
 reprint,
 amsmath,amssymb,
 aps,
]{revtex4-2}

\usepackage{graphicx}
\usepackage{dcolumn}
\usepackage{bm}

\begin{document}

\preprint{APS/123-QED}

\title{Phase Shift in AC Magnetocaloric Effect Measurements as an Indicator of the Order of Magnetic Phase Transitions}

\author{Akhmed M. Aliev}
\email{lowtemp@mail.ru}
\affiliation{%
 Amirkhanov Institute of Physics of Dagestan Federal Research Centre of RAS, 367003 Makhachkala, Russia}

\author{Adler G. Gamzatov}%
 \email{gamzatov_adler@mail.ru}
\affiliation{%
 Amirkhanov Institute of Physics of Dagestan Federal Research Centre of RAS, 367003 Makhachkala, Russia}

\author{Zaur Z. Alisultanov}%
\email{zaur0102@gmail.com}
 \affiliation{%
 Abrikosov Center for Theoretical Physics, MIPT, Dolgoprudnyi, Moscow Region, 141701, Russia
 }%
\affiliation{%
 Amirkhanov Institute of Physics of DFRC RAS, 367003 Makhachkala, Russia
 }%

\date{\today}

\begin{abstract}
It is shown that the phase shift between an applied weak alternating magnetic field and the magnetocaloric response signal of the magnetic material is drastically sensitive to the order of phase transition. Namely, at the second-order phase transition, the phase shift does not depend on the magnetic field magnitude, while in the first-order phase transition this one depends significantly on the field strength. We have shown that this effect follows from the general critical dynamics theory. 

\end{abstract}

\maketitle


According to the Ehrenfest's classification, the first derivative of the free energy with respect to some thermodynamic variable exhibits a discontinuity across phase transitions of the first order (FOPT). On the other hand, the second-order phase transitions (SOPT) are continuous in the first derivative but exhibit discontinuity in a second derivative of the free energy~\cite{LandauLifshitz,Jaeger,Ehrenfest,Sauer}. Behind each abstract derivative of the Gibbs thermodynamic potential there is a well-measured macroscopic parameter of the material. Accordingly, the order of phase transition can be determined by studying the temperature or pressure dependences of the corresponding parameters. 

One can also determine the type of phase transition according the behavior of the magnetocaloric effect (MCE). Materials with FOPT are characterized by temperature hysteresis of the magnetocaloric effect, one-sided temperature boundaries of the effect independent of the magnetic field (temperature effect limited from below or above), a non-monotonic dependence of the effect on the magnetic field, high asymmetry of the MCE curve with respect to temperature, etc.~\cite{Tishin2003,Aliev2016,Oliveira,Basso,Valiev}. But it needs to be said that sometimes establishing the phase transition order using MCE data are subjective, many of these criteria are not strictly defined. Often, strong magnetic fields are required to determine the order of phase transitions using MCE measurements. A strict quantitative parameter has also been proposed by which one can distinguish between transitions of the first and second order - the exponent n from the field dependence of the magnetic entropy change $\Delta S_M\sim H^n$ has a maximum of $n>2$ only for thermomagnetic FOPT~\cite{Law}. 

\begin{figure*}
\includegraphics[width=1.4\columnwidth]{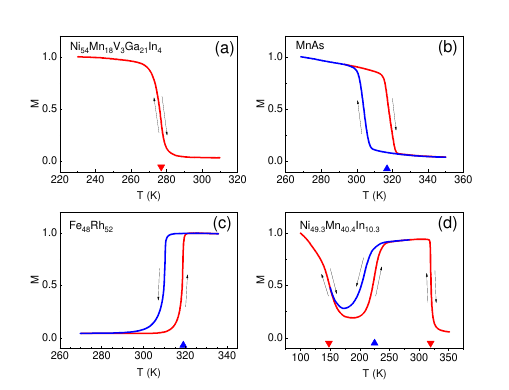}
\caption{\label{figure 1}Normalized temperature dependences of magnetization in a magnetic field of 200 Oe. Blue up triangles mark temperatures of FOPT (at heating run), red down triangles mark SOPT temperatures.}
\end{figure*}

Currently, MCE is studied using various methods~\cite{Tishin2003}. The most widely used method for indirectly estimating is evaluating the isothermal entropy change from magnetization and heat capacity data~\cite{Pecharsky,Amaral}. A direct technique for studying MCE is to measure the adiabatic temperature change of a material under a fast change in the applied external magnetic field~\cite{Tishin2003}. Currently, MCE is also being studied using theoretical methods~\cite{Sokolovsky}. When used in magnetic cooling technology, the magnetocaloric material will be exposed to a cyclic (alternating) magnetic field; accordingly, several techniques for measuring the MCE in alternating magnetic fields has been developed to carry out relevant studies~\cite{Gopal,Döntgen,Tokiwa,Aliev2010}. The essence of the method proposed elsewhere~\cite{Aliev2010} is that the magnetocaloric material is exposed to an alternating magnetic field; in general, such a field can be represented as
\begin{equation}
H=H_0 \text{sin}(\omega t)
\end{equation}
where $H_0$ is the amplitude value of the magnetic field, $\omega$ is the cyclic frequency of the magnetic field. The temperature response of a material to the applied alternating magnetic field in general can be presented in the form

\begin{equation}
\Delta T_{ad}=\pm\Delta T_{0} |\text{sin}(\omega t-\phi)|
\end{equation}
where $\Delta T_0$ is the amplitude value of the temperature change, $\phi$ is the phase shift between the magnetic field and the sample response. Phase shift occurs due to relaxation phenomena occurring during phase transitions. Due to the fact that the temperature response of different materials will be different because of the different field dependence of the MCE, it is not at all necessary that the temperature response will be functionally the same as the magnetic field. For MCE measurements this does not matter, the main condition is that this response occurs at the same frequency, which is the same as the disturbance frequency, and are in a certain way related in phase. The sign of $\Delta T$ will be positive in the case of direct and negative in the case of inverse MCE. The modulus of the function $\text{sin}(\omega t-\phi)$ means that we have $\Delta T$ of the certain sign, regardless of the direction of the magnetic field (in the absence of anisotropy). In the conventional direct MCE measurement, the measured parameter is only the adiabatic temperature change $\Delta T_{ad}$, and when measured in an alternating field, we have the values of two parameters - $\Delta T_{ad}$ and the phase shift $\phi$. A phase shift can provide valuable information about the behavior of a magnetic system in alternating magnetic fields, including some characteristics of magnetic phase transitions. For this purpose, we studied the magnetocaloric properties of several magnetocaloric materials with phase transitions of different nature, in alternating magnetic fields of low frequency and amplitude.

\begin{figure*}
\includegraphics[width=1.4\columnwidth]{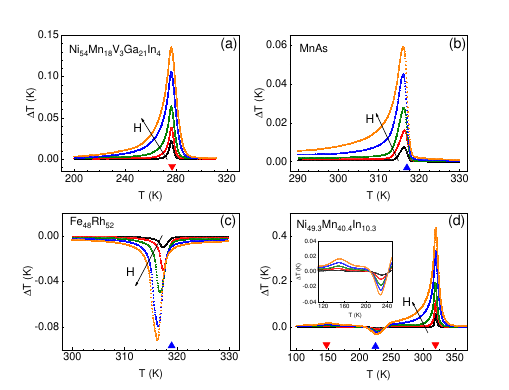}
\caption{\label{figure 2}Temperature dependences of the adiabatic temperature change $\Delta T_{ad}$ in alternating magnetic fields with an amplitude from 200, 500, 1000, 2000 and 3000 Oe. Blue up triangles mark temperatures of FOPT, red down triangles mark temperatures of SOPT (according to magnetization).}
\end{figure*}

The measurements were carried out in alternating magnetic fields according to the technique described in~\cite{Aliev2010}. The source of the alternating magnetic field was an electromagnet, through the coil of which an alternating current was passed using a current source with external analog control. To obtain alternating current, the external control input of the current source was supplied with voltage from the built-in AC generator of the Lock-in. Due to the inductance of the electromagnet coil, a temperature independent constant phase shift occurs between the magnetic field on the coil and the original control voltage. An additional constant phase shift may also appear between the AC voltage generator and current source. For our studies, such constant phase shifts do not matter, since the initial constant phase shift can always be set equal to zero using the phase shifter of the Lock-In.

We studied the magnetocaloric properties of several samples of different classes of magnetic materials: $\text{Ni}_{54}\text{Mn}_{18}\text{V}_{3}\text{Ga}_{21}\text{In}_{4}$, $\text{Ni}_{49.3}\text{Mn}_{40.4}\text{In}_{10.3}$ (Heusler alloys), $\text{Fe}_{48}\text{Rh}_{52}$ and $\text{MnAs}$. The choice of these materials is conditioned to that the most possible magnetic phase transitions can be observed in these materials. The magnetocaloric properties of these materials in moderate and high magnetic fields were previously studied elsewhere~\cite{Aliev2016,Aliev2020,Aliev2021}.

Fig.~\ref{figure 1} shows the normalized temperature dependences of the magnetization of all samples in a magnetic field of 200 Oe. In the $\text{Ni}_{54} \text{Mn}_{18} \text{V}_3 \text{Ga}_{21} \text{In}_4$ Heusler alloy, a second order ferromagnetic-paramagnetic phase transition occurs at the Curie temperature $T_C = 277 K$. In the $MnAs$ compound, a first-order ferromagnetic-paramagnetic phase transition is observed, with the Curie temperature $T_C = 317 K$ in the heating run and 304 K in the cooling run. In the $\text{Fe}_{48}\text{Rh}_{52}$ alloy, a first-order phase transition is observed from a low-temperature antiferromagnetic phase to a high-temperature ferromagnetic one, with a Néel temperatures $T_N = 319 K$ in heating run and 310 K in cooling run. In the $\text{Ni}_{49.3} \text{Mn}_{40.4} \text{In}_{10.3}$ Heusler alloy a series of phase transitions is observed: high-temperature second order ferromagnetic austenite-paramagnetic austenite phase transition with a Curie temperature of 319.5 K, low-temperature magnetostructural martensite - austenite phase transition, with Néel temperatures $T_N = 224 K$ in the heating run and 210 K in the cooling one. Further cooling in martensitic phase the sample transforms into a ferromagnetic state with the Curie temperature $T_C = 149 K$. All magnetostructural phase transitions are accompanied by a sharp change in lattice parameters. 

Fig.~\ref{figure 2} shows the temperature dependences of the adiabatic temperature change $\Delta T_{ad}$ of all samples in alternating magnetic fields 200÷3000 Oe. For clarity, the figures show the MCE curves only in heating runs. The results obtained reveal direct MCE in the case of FM-PM transitions, and inverse one in the region of the Néel temperature and in the region of the magnetostructural phase transition martensite-austenite. All materials studied reveals significant and even giant MCE values under moderate and high fields, but under weak field change the MCE values are small, this is especially typical in the region of the first-order phase transitions. This is a consequence of the fact that such fields are insufficient to induce magnetostructural transitions, and accordingly, there is no lattice contribution to the overall MCE. 

\begin{figure*}
\includegraphics[width=1.4\columnwidth]{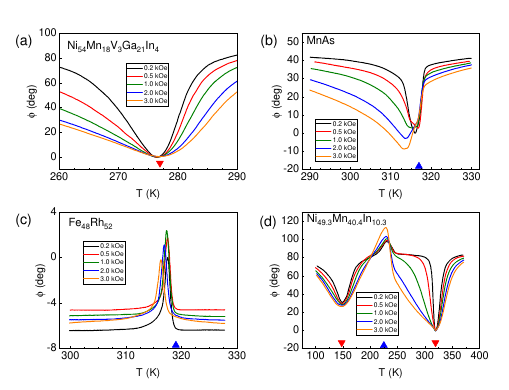}
\caption{\label{figure 3}Temperature dependences of the phase shift in alternating magnetic fields with an amplitude from 200, 500, 1000, 2000 and 3000 Oe. Blue up triangles mark temperatures of FOPT, red down triangles mark temperatures of SOPT (according to magnetization).}
\end{figure*}

The temperature dependences of the phase shift are shown in Fig.~\ref{figure 3}. When measuring the phase shift, the following procedure was carried out. Initially, $\Delta T_{ad}$ and the phase shift were measured in a magnetic field of 200 Oe. After that, using the phase shifter of the SR830 Lock-in, the phase was set to be zero at the phase transition point. In the case of the $\text{Ni}_{49.3} \text{Mn}_{40.4} \text{In}_{10.3}$ Heusler alloy, where several phase transitions are observed, the zero phase was set at the point of the ferromagnetic-paramagnetic phase transition (319.5 K). Then, with this phase set, measurements of $\Delta T_{ad}$ and phase shift were carried out in magnetic fields of 200, 500, 1000, 2000 and 3000 Oe.

The following peculiarities of the temperature dependences of the phase shift are observed in the Fig.~\ref{figure 3}. Firstly, in a narrow region near SOPT point, the phase shift weakly depends on the magnetic field. With distance from the Curie point, the phase shift changes and strongly depends on the magnetic field. As well, the phase shift in the region of the FOPTs strongly depends on the magnetic field. The points of maximum phase shifts at first order phase transitions shift in temperature with field, which is not observed at the point of second-order phase transitions.

Now we provide a qualitative interpretation of the above experimental data. The main effect is that at the point of a second-order phase transition the phase shift does not depend on the applied magnetic field. From the point of view of time dynamics, this means, in fact, that this shift does not depend on the field sweep rate. This follows from the fact that the field amplitude changes, but the frequency remains constant, i.e., the speed of magnetic field sweep changes. In other words, we are dealing with a situation where the response of the system (in this case, the adiabatic temperature change) at the transition point does not depend on the frequency of the disturbance. It turns out that such an absence of frequency dispersion at the second-order phase transition point can be described within the framework of the general theory of phase transitions and the theory of linear response.

We use the simplest fluctuation-dissipation model of phase transition. The main statement on which our qualitative interpretation is based is that at the point of a second-order phase transition fluctuations of the order parameter are static, i.e., do not depend on time, while at the point of a first-order phase transition, fluctuations depend on time, for example, they quickly decay. This statement follows from the theory of critical dynamics~\cite{Patashinskii,Hohenberg}. Let the order parameter be a scalar field
\begin{equation}
\Phi (r,t)=\Phi_0+\Delta\Phi (r,t)
\end{equation}
where $\Phi_0$ is the regular part of the order parameter, depending only on temperature, and $\Delta\Phi (r,t)$ are fluctuations of the order parameter, due to of which the order parameter can be considered as a field. If we introduce free energy as a functional of such a field $\mathcal{F}\left\{ \Phi\right\}$, then the rate of change $\Phi (r,t)$ for small deviations from equilibrium is proportional to the thermodynamic force~\cite{Patashinskii}
\begin{equation}
\partial_{t}\Phi\left(r,t\right)\equiv\partial_{t}\Delta\Phi\left(r,t\right)=-\Gamma\frac{\delta\mathcal{F}}{\delta\Phi}
\end{equation}
where $\Gamma$ is a kinetic coefficient that is finite at the phase transition point. From this expression it follows that at the point of the SOPT $\partial_{t}\Delta\Phi\left(r,t\right)=0$, because at this point $\delta\mathcal{F}/\delta\Phi=0$. Thus, at the transition point of the SOPT, fluctuations of the order parameter are static. Based on this property, we study the time dispersion of the thermodynamic quantity, i.e., the phase difference between the disturbance (in the case of MCE this is an alternating magnetic field) and the response (in the case of MCE this is the thermocouple signal caused by the magnetocaloric effect).

In general, the response $x(t)$  is related to the perturbation $f(t)$  using the relation (within the framework of linear response theory) 
\begin{equation}
x\left(t\right)=\int\alpha\left(t-\tau\right)f\left(\tau\right)d\tau
\end{equation}
where $\alpha\left(t-\tau\right)$ is the response function (susceptibility). In the case of MCE, the quantity $x\left(t\right)$ is the magnetization, and the perturbation is the magnetic field. The Fourier transform gives $\overline{x}\left(\omega\right)=\overline{\alpha}\left(\omega\right)\overline{f}\left(\omega\right)$, where the symbols with an overbar denote the Fourier images of the corresponding functions. In the general case $\overline{\alpha}\left(\omega\right)=\overline{\alpha}'\left(\omega\right)+i\overline{\alpha}''\left(\omega\right)$. The presence of a frequency dependence in the real part of the susceptibility $\overline{\alpha}'\left(\omega\right)$ means the presence of frequency or time dispersion in the system, due to which the response function lags compared to the disturbance function.

The fluctuation-dissipation theorem (FDT) relates the imaginary part of susceptibility with the so-called spectral density of fluctuations $\overline{S}_{x}\left(\omega\right)$ of the thermodynamic quantity $x\left(t\right)$
\begin{equation}
\overline{S}_{x}\left(\omega\right)=\hbar\overline{\alpha}''\left(\omega\right)\coth\left(\frac{\hbar\omega}{kT}\right)
\end{equation}
In the limit of low frequencies, when $\hbar\omega\ll kT$ (classical limit) we have
\begin{equation}
\overline{\alpha}''\left(\omega\right)\approx\frac{\omega}{2kT}\overline{S}_{x}\left(\omega\right)
\end{equation}

Now let us take advantage of the fact that at the point of a second-order phase transition the fluctuations are static. This means that the value of $S_{x}$ at the transition point does not depend on time: $S_{x}=const$ (near SOPT the time evolution of this function is $S_{x}\sim\exp\left(-t/t_{\xi}\right)$, where $t_{\xi}=\xi^{\kappa}$ and $\xi$ is the fluctuation size, $\kappa$ is some positive number), which gives for their Fourier transform (spectral density)
\begin{equation}
\overline{S}_{x}\left(\omega\right)=\text{const}2\pi\delta\left(\omega\right)
\end{equation}
Thus, for a SOPT we obtain
\begin{equation}
\overline{\alpha}''\left(\omega\right)\approx\frac{\pi\omega}{kT}\text{const}\delta\left(\omega\right)
\end{equation}
Let's find the real part of the susceptibility using the Kramers-Kronig relations $\overline{\alpha}'\left(\omega\right)=\frac{1}{\pi}P\int\frac{\overline{\alpha}''\left(z\right)dz}{z-\omega}$. This integral is nonzero only at $\omega=0$. Thus
\begin{equation}
\overline{\alpha}'\left(\omega\right)=\frac{\text{const}}{kT} \delta_{\omega,0}
\end{equation}

Thus, at the SOPT point there is no dispersion. At this point there should be no lag in the phase of the response from the disturbance, i.e., $\phi=0$. For FOPTs fluctuations have a short lifetime, i.e., not static. This means that the $\overline{\alpha}'\left(\omega\right)$ value will be a function of frequency. So, we can say that the FOPT point is characterized by the presence of susceptibility dispersion, and therefore a phase shift. This shift is measured by a Lock-in. If the Lock-in input receives an MCE signal in the form $u_i=\Delta T_{0}\text{sin}(\omega t+\phi)$, then the output signal turns out to be proportional to the amplitude and also contains a phase factor: $u_f=\Delta T_{0}\text{cos}\phi$. From this we get $\varphi=\arccos\left(\frac{u_{f}}{\Delta T_{0}}\right)$ . Because the temperature change $\Delta T_{0}$ during MCE is a nonlinear function of the magnetic field (this can be understood from Maxwell’s relations, for instance), then the $\phi$ will be a function of the magnetic field amplitude. Let's demonstrate this for a simple case when the fluctuation correlator has the form $S_{x}=\text{const}\exp\left(-\gamma |t|\right)$, where $\gamma$ is the inverse fluctuation lifetime (for second-order transitions $\gamma=0$). In this case
\begin{equation}
\overline{\alpha}'\left(\omega\right)=\frac{1}{kT}\left[\frac{\gamma^{2}}{\omega^{2}+\gamma^{2}}\left(1-\delta_{\omega,0}\right)+\delta_{\omega,0}\right]
\end{equation}
Applying the inverse Fourier transform, we obtain
\begin{equation}
\alpha'\left(t\right)=\frac{1}{kT}\left(\pi\gamma e^{-\gamma\left|t\right|}+\delta_{\gamma,0}\right)
\end{equation}
So, we have obtained an expression for the real part of the susceptibility for the case of fluctuations with a finite lifetime. This function describes the time dispersion in the system, i.e., delay of the response signal from the perturbation. Let's show this directly. Let the perturbation is given in the form $f\left(t\right)=\cos\omega t$. Then for the response we obtain
\begin{equation}
x\left(t\right)\sim\text{Re}\frac{e^{i\omega t}-e^{-\gamma t}}{i\omega+\gamma}+\delta_{\gamma,0}\sin\omega t
\end{equation}
The quantity $\gamma$ is large for FOPT. Therefore, we can neglect the term $e^{-\gamma t}$. Then we have (at $\gamma \neq 0$)
\begin{equation}
x\left(t\right)=\frac{\pi\gamma}{kT}\frac{\gamma\cos\omega t+\omega\sin\omega t}{\omega^{2}+\gamma^{2}}
\end{equation}
If we put $\gamma=A \text{sin}\phi$ and $\omega=A \text{cos}\phi$, then
\begin{equation}
x\left(t\right)\sim\frac{\pi\gamma}{kT\sqrt{\omega^{2}+\gamma^{2}}}\sin\left(\omega t+\phi\right)
\end{equation}
Thus, the finite lifetime of fluctuations leads to a phase shift $\phi=\arctan\left(\gamma/\omega\right)$ of the response from the perturbation. From this it can be seen that at $\gamma=0$ (SOPT) the phase shift is zero. The presented qualitative picture of the time dynamics is completely consistent with the experimental data.

From the results obtained, it can be stated that in the immediate vicinity of magnetic phase transitions there is no dependence of the phase shift between the magnetic field and the temperature response of the sample on the magnetic field and the rate of change of the magnetic field. Thus, this parameter can be used as an indicator of the order of phase transition. The absence of a dependence of the phase shift on the field indicates a second-order phase transition, and the dependence on the field of the phase shift and the temperature shift of the point of maximum phase shift indicate a first-order phase transition.

\begin{acknowledgments}
The work is supported by the Russian Science Foundation Grant No. 22-19-00610
\end{acknowledgments}

\bibliography{apssamp}

\providecommand{\noopsort}[1]{}\providecommand{\singleletter}[1]{#1}%
\begin{thebibliography}{21}%
\makeatletter
\providecommand \@ifxundefined [1]{%
 \@ifx{#1\undefined}
}%
\providecommand \@ifnum [1]{%
 \ifnum #1\expandafter \@firstoftwo
 \else \expandafter \@secondoftwo
 \fi
}%
\providecommand \@ifx [1]{%
 \ifx #1\expandafter \@firstoftwo
 \else \expandafter \@secondoftwo
 \fi
}%
\providecommand \natexlab [1]{#1}%
\providecommand \enquote  [1]{``#1''}%
\providecommand \bibnamefont  [1]{#1}%
\providecommand \bibfnamefont [1]{#1}%
\providecommand \citenamefont [1]{#1}%
\providecommand \href@noop [0]{\@secondoftwo}%
\providecommand \href [0]{\begingroup \@sanitize@url \@href}%
\providecommand \@href[1]{\@@startlink{#1}\@@href}%
\providecommand \@@href[1]{\endgroup#1\@@endlink}%
\providecommand \@sanitize@url [0]{\catcode `\\12\catcode `\$12\catcode `\&12\catcode `\#12\catcode `\^12\catcode `\_12\catcode `\%12\relax}%
\providecommand \@@startlink[1]{}%
\providecommand \@@endlink[0]{}%
\providecommand \url  [0]{\begingroup\@sanitize@url \@url }%
\providecommand \@url [1]{\endgroup\@href {#1}{\urlprefix }}%
\providecommand \urlprefix  [0]{URL }%
\providecommand \Eprint [0]{\href }%
\providecommand \doibase [0]{https://doi.org/}%
\providecommand \selectlanguage [0]{\@gobble}%
\providecommand \bibinfo  [0]{\@secondoftwo}%
\providecommand \bibfield  [0]{\@secondoftwo}%
\providecommand \translation [1]{[#1]}%
\providecommand \BibitemOpen [0]{}%
\providecommand \bibitemStop [0]{}%
\providecommand \bibitemNoStop [0]{.\EOS\space}%
\providecommand \EOS [0]{\spacefactor3000\relax}%
\providecommand \BibitemShut  [1]{\csname bibitem#1\endcsname}%
\let\auto@bib@innerbib\@empty
\bibitem [{\citenamefont {Landau}\ and\ \citenamefont {Lifshitz}(1994)}]{LandauLifshitz}%
  \BibitemOpen
  \bibfield  {author} {\bibinfo {author} {\bibfnamefont {.~L.}\ \bibnamefont {Landau}}\ and\ \bibinfo {author} {\bibfnamefont {E.}~\bibnamefont {Lifshitz}},\ }\href@noop {} {\emph {\bibinfo {title} {Statistical Physics Part 1, vol. 5 of Course of Theoretical Physics}}}\ (\bibinfo  {publisher} {Pergamon, 3rd Ed.},\ \bibinfo {year} {1994})\BibitemShut {NoStop}%
\bibitem [{\citenamefont {Jaeger}(1998)}]{Jaeger}%
  \BibitemOpen
  \bibfield  {author} {\bibinfo {author} {\bibfnamefont {G.}~\bibnamefont {Jaeger}},\ }\bibfield  {title} {\bibinfo {title} {The ehrenfest classification of phase transitions: Introduction and evolution},\ }\href@noop {} {\bibfield  {journal} {\bibinfo  {journal} {Archive for History of Exact Sciences}\ }\textbf {\bibinfo {volume} {53}},\ \bibinfo {pages} {51} (\bibinfo {year} {1998})}\BibitemShut {NoStop}%
\bibitem [{\citenamefont {Ehrenfest}(1933)}]{Ehrenfest}%
  \BibitemOpen
  \bibfield  {author} {\bibinfo {author} {\bibfnamefont {P.}~\bibnamefont {Ehrenfest}},\ }\bibfield  {title} {\bibinfo {title} {Phasenumwandlungen im ueblichen und erweiterten sinn, classifiziert nach dem entsprechenden singularitaeten des thermodynamischen potentiales},\ }\href@noop {} {\bibfield  {journal} {\bibinfo  {journal} {Verhandlingen der Koninklijke Akademie van Wetenschappen (Amsterdam)}\ }\textbf {\bibinfo {volume} {36}},\ \bibinfo {pages} {153–157} (\bibinfo {year} {1933})}\BibitemShut {NoStop}%
\bibitem [{\citenamefont {Sauer}(2017)}]{Sauer}%
  \BibitemOpen
  \bibfield  {author} {\bibinfo {author} {\bibfnamefont {T.}~\bibnamefont {Sauer}},\ }\bibfield  {title} {\bibinfo {title} {A look back at the ehrenfest classification translation and commentary of ehrenfest’s 1933 paper introducing the notion of phase transitions of different order},\ }\href@noop {} {\bibfield  {journal} {\bibinfo  {journal} {Verhandlingen der Koninklijke Akademie van Wetenschappen (Amsterdam)}\ }\textbf {\bibinfo {volume} {226}},\ \bibinfo {pages} {539–549} (\bibinfo {year} {2017})}\BibitemShut {NoStop}%
\bibitem [{\citenamefont {Tishin}\ and\ \citenamefont {Spichin}(2003)}]{Tishin2003}%
  \BibitemOpen
  \bibfield  {author} {\bibinfo {author} {\bibfnamefont {A.}~\bibnamefont {Tishin}}\ and\ \bibinfo {author} {\bibfnamefont {Y.}~\bibnamefont {Spichin}},\ }\href@noop {} {\emph {\bibinfo {title} {The Magnetocaloric Effect and its Applications}}}\ (\bibinfo  {publisher} {IOP Publishing, London},\ \bibinfo {year} {2003})\BibitemShut {NoStop}%
\bibitem [{\citenamefont {Aliev}\ \emph {et~al.}(2016)\citenamefont {Aliev}, \citenamefont {Batdalov}, \citenamefont {Khanov}, \citenamefont {Kamantsev}, \citenamefont {Koledov}, \citenamefont {Mashirov}, \citenamefont {Shavrov}, \citenamefont {Grechishkin}, \citenamefont {Kaul},\ and\ \citenamefont {Sampath}}]{Aliev2016}%
  \BibitemOpen
  \bibfield  {author} {\bibinfo {author} {\bibfnamefont {A.~M.}\ \bibnamefont {Aliev}}, \bibinfo {author} {\bibfnamefont {A.~B.}\ \bibnamefont {Batdalov}}, \bibinfo {author} {\bibfnamefont {L.~N.}\ \bibnamefont {Khanov}}, \bibinfo {author} {\bibfnamefont {A.~P.}\ \bibnamefont {Kamantsev}}, \bibinfo {author} {\bibfnamefont {V.~V.}\ \bibnamefont {Koledov}}, \bibinfo {author} {\bibfnamefont {A.~V.}\ \bibnamefont {Mashirov}}, \bibinfo {author} {\bibfnamefont {V.~G.}\ \bibnamefont {Shavrov}}, \bibinfo {author} {\bibfnamefont {R.~M.}\ \bibnamefont {Grechishkin}}, \bibinfo {author} {\bibfnamefont {A.~R.}\ \bibnamefont {Kaul}},\ and\ \bibinfo {author} {\bibfnamefont {V.}~\bibnamefont {Sampath}},\ }\bibfield  {title} {\bibinfo {title} {Reversible magnetocaloric effect in materials with first order phase transitions in cyclic magnetic fields: Fe48rh52 and sm0.6sr0.4mno3},\ }\href@noop {} {\bibfield  {journal} {\bibinfo  {journal} {Appl. Phys. Lett.}\ }\textbf {\bibinfo {volume} {109}},\ \bibinfo {pages} {202407}
  (\bibinfo {year} {2016})}\BibitemShut {NoStop}%
\bibitem [{\citenamefont {de~Oliveira}\ and\ \citenamefont {von Ranke}(2008)}]{Oliveira}%
  \BibitemOpen
  \bibfield  {author} {\bibinfo {author} {\bibfnamefont {N.~A.}\ \bibnamefont {de~Oliveira}}\ and\ \bibinfo {author} {\bibfnamefont {P.~J.}\ \bibnamefont {von Ranke}},\ }\bibfield  {title} {\bibinfo {title} {Magnetocaloric effect around a magnetic phase transition},\ }\href@noop {} {\bibfield  {journal} {\bibinfo  {journal} {Phys. Rev. B}\ }\textbf {\bibinfo {volume} {77}},\ \bibinfo {pages} {214439} (\bibinfo {year} {2008})}\BibitemShut {NoStop}%
\bibitem [{\citenamefont {Basso}(2011)}]{Basso}%
  \BibitemOpen
  \bibfield  {author} {\bibinfo {author} {\bibfnamefont {V.}~\bibnamefont {Basso}},\ }\bibfield  {title} {\bibinfo {title} {The magnetocaloric effect at the first-order magneto-elastic phase transition},\ }\href@noop {} {\bibfield  {journal} {\bibinfo  {journal} {J. Phys.: Condens. Matter}\ }\textbf {\bibinfo {volume} {23}},\ \bibinfo {pages} {226004} (\bibinfo {year} {2011})}\BibitemShut {NoStop}%
\bibitem [{\citenamefont {Valiev}(2009)}]{Valiev}%
  \BibitemOpen
  \bibfield  {author} {\bibinfo {author} {\bibfnamefont {E.}~\bibnamefont {Valiev}},\ }\bibfield  {title} {\bibinfo {title} {Entropy and magnetocaloric effects in ferromagnets undergoing first- and second-order magnetic phase transitions},\ }\href@noop {} {\bibfield  {journal} {\bibinfo  {journal} {Journal of Experimental and Theoretical Physics}\ }\textbf {\bibinfo {volume} {108}},\ \bibinfo {pages} {279–285} (\bibinfo {year} {2009})}\BibitemShut {NoStop}%
\bibitem [{\citenamefont {Law}\ \emph {et~al.}(2018)\citenamefont {Law}, \citenamefont {Franco}, \citenamefont {Moreno-Ramírez}, \citenamefont {Conde}, \citenamefont {Karpenkov}, \citenamefont {Radulov}, \citenamefont {Skokov},\ and\ \citenamefont {Gutfleisch}}]{Law}%
  \BibitemOpen
  \bibfield  {author} {\bibinfo {author} {\bibfnamefont {J.~Y.}\ \bibnamefont {Law}}, \bibinfo {author} {\bibfnamefont {V.}~\bibnamefont {Franco}}, \bibinfo {author} {\bibfnamefont {L.~M.}\ \bibnamefont {Moreno-Ramírez}}, \bibinfo {author} {\bibfnamefont {A.}~\bibnamefont {Conde}}, \bibinfo {author} {\bibfnamefont {D.~Y.}\ \bibnamefont {Karpenkov}}, \bibinfo {author} {\bibfnamefont {I.}~\bibnamefont {Radulov}}, \bibinfo {author} {\bibfnamefont {K.~P.}\ \bibnamefont {Skokov}},\ and\ \bibinfo {author} {\bibfnamefont {O.}~\bibnamefont {Gutfleisch}},\ }\bibfield  {title} {\bibinfo {title} {A quantitative criterion for determining the order of magnetic phase transitions using the magnetocaloric effect},\ }\href@noop {} {\bibfield  {journal} {\bibinfo  {journal} {Nature Communications}\ }\textbf {\bibinfo {volume} {9}},\ \bibinfo {pages} {2680} (\bibinfo {year} {2018})}\BibitemShut {NoStop}%
\bibitem [{\citenamefont {Pecharsky}\ and\ \citenamefont {Gschneidner}(1999)}]{Pecharsky}%
  \BibitemOpen
  \bibfield  {author} {\bibinfo {author} {\bibfnamefont {V.~K.}\ \bibnamefont {Pecharsky}}\ and\ \bibinfo {author} {\bibfnamefont {K.~A.}\ \bibnamefont {Gschneidner}},\ }\bibfield  {title} {\bibinfo {title} {Magnetocaloric effect from indirect measurements: Magnetization and heat capacity},\ }\href@noop {} {\bibfield  {journal} {\bibinfo  {journal} {J. Appl. Phys.}\ }\textbf {\bibinfo {volume} {86}},\ \bibinfo {pages} {565} (\bibinfo {year} {1999})}\BibitemShut {NoStop}%
\bibitem [{\citenamefont {Amaral}\ and\ \citenamefont {Amaral}(2010)}]{Amaral}%
  \BibitemOpen
  \bibfield  {author} {\bibinfo {author} {\bibfnamefont {J.~S.}\ \bibnamefont {Amaral}}\ and\ \bibinfo {author} {\bibfnamefont {V.~S.}\ \bibnamefont {Amaral}},\ }\bibfield  {title} {\bibinfo {title} {On estimating the magnetocaloric effect from magnetization measurements},\ }\href@noop {} {\bibfield  {journal} {\bibinfo  {journal} {Journal of Magnetism and Magnetic Materials}\ }\textbf {\bibinfo {volume} {322}},\ \bibinfo {pages} {1552} (\bibinfo {year} {2010})}\BibitemShut {NoStop}%
\bibitem [{\citenamefont {Sokolovsky}\ \emph {et~al.}(2022)\citenamefont {Sokolovsky}, \citenamefont {Miroshkina},\ and\ \citenamefont {Buchelnikov}}]{Sokolovsky}%
  \BibitemOpen
  \bibfield  {author} {\bibinfo {author} {\bibfnamefont {V.~V.}\ \bibnamefont {Sokolovsky}}, \bibinfo {author} {\bibfnamefont {O.~N.}\ \bibnamefont {Miroshkina}},\ and\ \bibinfo {author} {\bibfnamefont {V.~D.}\ \bibnamefont {Buchelnikov}},\ }\bibfield  {title} {\bibinfo {title} {Review of modern theoretical methods for studying magnetocaloric materials},\ }\href@noop {} {\bibfield  {journal} {\bibinfo  {journal} {Physics of Metals and Metallography}\ }\textbf {\bibinfo {volume} {123}},\ \bibinfo {pages} {344} (\bibinfo {year} {2022})}\BibitemShut {NoStop}%
\bibitem [{\citenamefont {Gopal}\ \emph {et~al.}(1995)\citenamefont {Gopal}, \citenamefont {Chahine}, \citenamefont {Foldeaki},\ and\ \citenamefont {Bose}}]{Gopal}%
  \BibitemOpen
  \bibfield  {author} {\bibinfo {author} {\bibfnamefont {B.~R.}\ \bibnamefont {Gopal}}, \bibinfo {author} {\bibfnamefont {R.}~\bibnamefont {Chahine}}, \bibinfo {author} {\bibfnamefont {M.~M.}\ \bibnamefont {Foldeaki}},\ and\ \bibinfo {author} {\bibfnamefont {T.~K.}\ \bibnamefont {Bose}},\ }\bibfield  {title} {\bibinfo {title} {Noncontact thermoacoustic method to measure the magnetocaloric effect},\ }\href@noop {} {\bibfield  {journal} {\bibinfo  {journal} {Rev. Sci. Instrum.}\ }\textbf {\bibinfo {volume} {66}},\ \bibinfo {pages} {232} (\bibinfo {year} {1995})}\BibitemShut {NoStop}%
\bibitem [{\citenamefont {Döntgen}\ \emph {et~al.}(2018)\citenamefont {Döntgen}, \citenamefont {Rudolph}, \citenamefont {Waske},\ and\ \citenamefont {Hägele}}]{Döntgen}%
  \BibitemOpen
  \bibfield  {author} {\bibinfo {author} {\bibfnamefont {J.}~\bibnamefont {Döntgen}}, \bibinfo {author} {\bibfnamefont {J.}~\bibnamefont {Rudolph}}, \bibinfo {author} {\bibfnamefont {A.}~\bibnamefont {Waske}},\ and\ \bibinfo {author} {\bibfnamefont {D.}~\bibnamefont {Hägele}},\ }\bibfield  {title} {\bibinfo {title} {Modulation infrared thermometry of caloric effects at up to khz frequencies},\ }\href@noop {} {\bibfield  {journal} {\bibinfo  {journal} {Rev. Sci. Instrum.}\ }\textbf {\bibinfo {volume} {89}},\ \bibinfo {pages} {033909} (\bibinfo {year} {2018})}\BibitemShut {NoStop}%
\bibitem [{\citenamefont {Tokiwa}\ and\ \citenamefont {Gegenwart}(2011)}]{Tokiwa}%
  \BibitemOpen
  \bibfield  {author} {\bibinfo {author} {\bibfnamefont {Y.}~\bibnamefont {Tokiwa}}\ and\ \bibinfo {author} {\bibfnamefont {P.}~\bibnamefont {Gegenwart}},\ }\bibfield  {title} {\bibinfo {title} {High-resolution alternating-field technique to determine the magnetocaloric effect of metals down to very low temperatures},\ }\href@noop {} {\bibfield  {journal} {\bibinfo  {journal} {Rev. Sci. Instrum.}\ }\textbf {\bibinfo {volume} {82}},\ \bibinfo {pages} {013905} (\bibinfo {year} {2011})}\BibitemShut {NoStop}%
\bibitem [{\citenamefont {Aliev}\ \emph {et~al.}(2010)\citenamefont {Aliev}, \citenamefont {Batdalov},\ and\ \citenamefont {Kalitka}}]{Aliev2010}%
  \BibitemOpen
  \bibfield  {author} {\bibinfo {author} {\bibfnamefont {A.~M.}\ \bibnamefont {Aliev}}, \bibinfo {author} {\bibfnamefont {A.~B.}\ \bibnamefont {Batdalov}},\ and\ \bibinfo {author} {\bibfnamefont {V.~S.}\ \bibnamefont {Kalitka}},\ }\bibfield  {title} {\bibinfo {title} {Magnetocaloric properties of manganites in alternating magnetic fields},\ }\href@noop {} {\bibfield  {journal} {\bibinfo  {journal} {JETP Letters}\ }\textbf {\bibinfo {volume} {90}},\ \bibinfo {pages} {663–666} (\bibinfo {year} {2010})}\BibitemShut {NoStop}%
\bibitem [{\citenamefont {Aliev}\ \emph {et~al.}(2020)\citenamefont {Aliev}, \citenamefont {Batdalov}, \citenamefont {Khanov}, \citenamefont {Mashirov}, \citenamefont {Dil’mieva}, \citenamefont {Koledov},\ and\ \citenamefont {Shavrov}}]{Aliev2020}%
  \BibitemOpen
  \bibfield  {author} {\bibinfo {author} {\bibfnamefont {A.~M.}\ \bibnamefont {Aliev}}, \bibinfo {author} {\bibfnamefont {A.~B.}\ \bibnamefont {Batdalov}}, \bibinfo {author} {\bibfnamefont {L.~N.}\ \bibnamefont {Khanov}}, \bibinfo {author} {\bibfnamefont {A.~V.}\ \bibnamefont {Mashirov}}, \bibinfo {author} {\bibfnamefont {E.~T.}\ \bibnamefont {Dil’mieva}}, \bibinfo {author} {\bibfnamefont {V.~V.}\ \bibnamefont {Koledov}},\ and\ \bibinfo {author} {\bibfnamefont {V.~G.}\ \bibnamefont {Shavrov}},\ }\bibfield  {title} {\bibinfo {title} {Degradation of the magnetocaloric effect in ni49.3mn40.4in10.3 in a cyclic magnetic field},\ }\href@noop {} {\bibfield  {journal} {\bibinfo  {journal} {Physics of the Solid State}\ }\textbf {\bibinfo {volume} {62}},\ \bibinfo {pages} {837–840} (\bibinfo {year} {2020})}\BibitemShut {NoStop}%
\bibitem [{\citenamefont {Aliev}\ \emph {et~al.}(2021)\citenamefont {Aliev}, \citenamefont {Khanov}, \citenamefont {Gamzatov}, \citenamefont {Batdalov}, \citenamefont {Kurbanova}, \citenamefont {Yanushkevich},\ and\ \citenamefont {Govor}}]{Aliev2021}%
  \BibitemOpen
  \bibfield  {author} {\bibinfo {author} {\bibfnamefont {A.~M.}\ \bibnamefont {Aliev}}, \bibinfo {author} {\bibfnamefont {L.~N.}\ \bibnamefont {Khanov}}, \bibinfo {author} {\bibfnamefont {A.~G.}\ \bibnamefont {Gamzatov}}, \bibinfo {author} {\bibfnamefont {A.~B.}\ \bibnamefont {Batdalov}}, \bibinfo {author} {\bibfnamefont {D.~R.}\ \bibnamefont {Kurbanova}}, \bibinfo {author} {\bibfnamefont {K.~I.}\ \bibnamefont {Yanushkevich}},\ and\ \bibinfo {author} {\bibfnamefont {G.~A.}\ \bibnamefont {Govor}},\ }\bibfield  {title} {\bibinfo {title} {Giant magnetocaloric effect in mnas1-xpx in a cyclic magnetic field: Lattice and magnetic contributions and degradation of the effect},\ }\href@noop {} {\bibfield  {journal} {\bibinfo  {journal} {Appl. Phys. Lett.}\ }\textbf {\bibinfo {volume} {118}},\ \bibinfo {pages} {072404} (\bibinfo {year} {2021})}\BibitemShut {NoStop}%
\bibitem [{\citenamefont {Patashinskii}\ and\ \citenamefont {Pokrovskii}(1979)}]{Patashinskii}%
  \BibitemOpen
  \bibfield  {author} {\bibinfo {author} {\bibfnamefont {A.~Z.}\ \bibnamefont {Patashinskii}}\ and\ \bibinfo {author} {\bibfnamefont {V.~L.}\ \bibnamefont {Pokrovskii}},\ }\href@noop {} {\emph {\bibinfo {title} {Fluctuation Theory of Phase Transitions}}}\ (\bibinfo  {publisher} {Pergamon Press},\ \bibinfo {year} {1979})\BibitemShut {NoStop}%
\bibitem [{\citenamefont {Hohenberg}\ and\ \citenamefont {Halperin}(1977)}]{Hohenberg}%
  \BibitemOpen
  \bibfield  {author} {\bibinfo {author} {\bibfnamefont {P.~C.}\ \bibnamefont {Hohenberg}}\ and\ \bibinfo {author} {\bibfnamefont {B.~I.}\ \bibnamefont {Halperin}},\ }\bibfield  {title} {\bibinfo {title} {Theory of dynamic critical phenomena},\ }\href@noop {} {\bibfield  {journal} {\bibinfo  {journal} {Rev. Mod. Phys.}\ }\textbf {\bibinfo {volume} {49}},\ \bibinfo {pages} {435} (\bibinfo {year} {1977})}\BibitemShut {NoStop}%
\end{thebibliography}%

\end{document}